\documentclass[conference]{IEEEtran}
\IEEEoverridecommandlockouts
\usepackage{cite}
\usepackage{amsmath,amssymb,amsfonts}
\usepackage{algorithmic}
\usepackage{graphicx}
\usepackage{textcomp}
\usepackage[pscoord]{eso-pic}
\usepackage[a4paper, total={184mm,239mm}]{geometry}
\usepackage{xcolor}
\usepackage{balance}
\usepackage{subcaption}
\usepackage{amsmath}
\usepackage{tabularx}
\usepackage{authblk}
\def\BibTeX{{\rm B\kern-.05em{\sc i\kern-.025em b}\kern-.08em
    T\kern-.1667em\lower.7ex\hbox{E}\kern-.125emX}}
\newcommand{\magenta}[1]{\textcolor{black}{#1}}

\newcommand{\blue}[1]{\textcolor{black}{#1}}
\newcommand{\cyan}[1]{\textcolor{black}{#1}}
\newcommand{\orange}[1]{\textcolor{black}{#1}}
\newcommand{\pink}[1]{\textcolor{black}{#1}}

\newcommand{\unet}{{\textit{U-Net}}}
\newcommand{\technologyname}{{\textit{Mono3D}}}
\newcommand{\resnet}{{\textit{ResNet-50}}}
\begin{document}
\title{Temperature-Aware Monolithic 3D DNN Accelerators for Biomedical Applications
}
\author[1]{\normalsize Prachi Shukla}
\author[2]{Vasilis F. Pavlidis}
\author[3]{Emre Salman}
\author[1]{Ayse K. Coskun}

\affil[1]{\normalsize Boston University, Boston, USA - \{prachis, acoskun\}@bu.edu}
\affil[2]{The University of Manchester, Manchester, UK - vasileios.pavlidis@manchester.ac.uk}
\affil[3]{Stony Brook University, Stony Brook, USA - emre.salman@stonybrook.edu }
\maketitle
\begin{abstract}

In this paper, we focus on temperature-aware Monolithic 3D (\technologyname{}) deep neural network (DNN) inference accelerators for biomedical applications. 
We \pink{develop} an optimizer that tunes aspect ratios and footprint of the accelerator under user-defined performance and thermal constraints, and generates near-optimal configurations. Using the \pink{proposed} \technologyname{} optimizer, we \pink{demonstrate} up to 61\% improvement in energy efficiency 
for biomedical applications over a performance-optimized accelerator.\let\thefootnote\relax\footnotetext{This paper was accepted to be presented at the Design, Automation and Test in Europe Conference (DATE) 2022 workshop on ``3D Integration: Heterogeneous 3D Architectures and Sensors".}
\end{abstract}


\section{Introduction}
\label{sec:introducton}
\blue{Deep neural network }(DNN) inference is widely used for image segmentation and recognition in biomedical applications, \magenta{e.g., improving 
\cyan{imaging} for cancer detection} \cite{ronneberger2015u,al2020breast}. For these applications, mobile/portable DNN accelerators are in demand to optimize for computation speed, energy efficiency, and small footprint
 \cite{wei2020review}. Monolithic 3D (\technologyname{}) is an emerging 3D \pink{technology} \cyan{with} the potential to offer these characteristics and \pink{provide improvement} 
over 2D systems \cite{batude20143d}.


In \technologyname{} \pink{ICs}, two or more thin tiers are vertically integrated in a sequential fabrication process, 
\cyan{\pink{where} nanometer-scale vias provide high-density vertical interconnects,}
\cyan{thus leading} to \magenta{dense} integration. Due to the thin tiers, \technologyname{} has lower vertical thermal resistance than other 3D technologies, 
\cyan{e.g., 3D stacking}
\cite{hu2018stacking}, and results in \pink{strong} inter-tier thermal coupling. 
Furthermore, the \pink{strong} thermal coupling may lead to \cyan{similar} high density hot \orange{spots 
across} tiers \cite{shukla2019overview}. \pink{In addition}, the absence of heat sinks and fans in mobile systems can \magenta{escalate} thermal concerns. Therefore, it is imperative to consider thermal awareness while designing mobile \technologyname{} systems. 

To provide energy and area efficiency, while also maintaining thermal integrity in \technologyname{} systems, we utilize an existing temperature-aware optimizer to generate near-optimal \blue{mobile} DNN accelerator configurations for biomedical applications. In this work, we use systolic arrays as the target DNN accelerator due to \magenta{their} simple architecture \pink{\cite{kung1982systolic}}. We investigate two DNNs (\unet{}, \resnet{}) that are used for image segmentation and classification, respectively, due to their high accuracy.

\section{Temperature-aware \technologyname{} Systolic Arrays}

\begin{figure}[tb]
	\centering
		\subfloat[\blue{\technologyname{} optimizer.}]{
	\vspace{-0.05in}
	\label{fig:optimizer}
	\includegraphics[trim= 0mm 0cm 0mm 0mm,clip,height=3.5cm,width=.8\columnwidth,keepaspectratio]{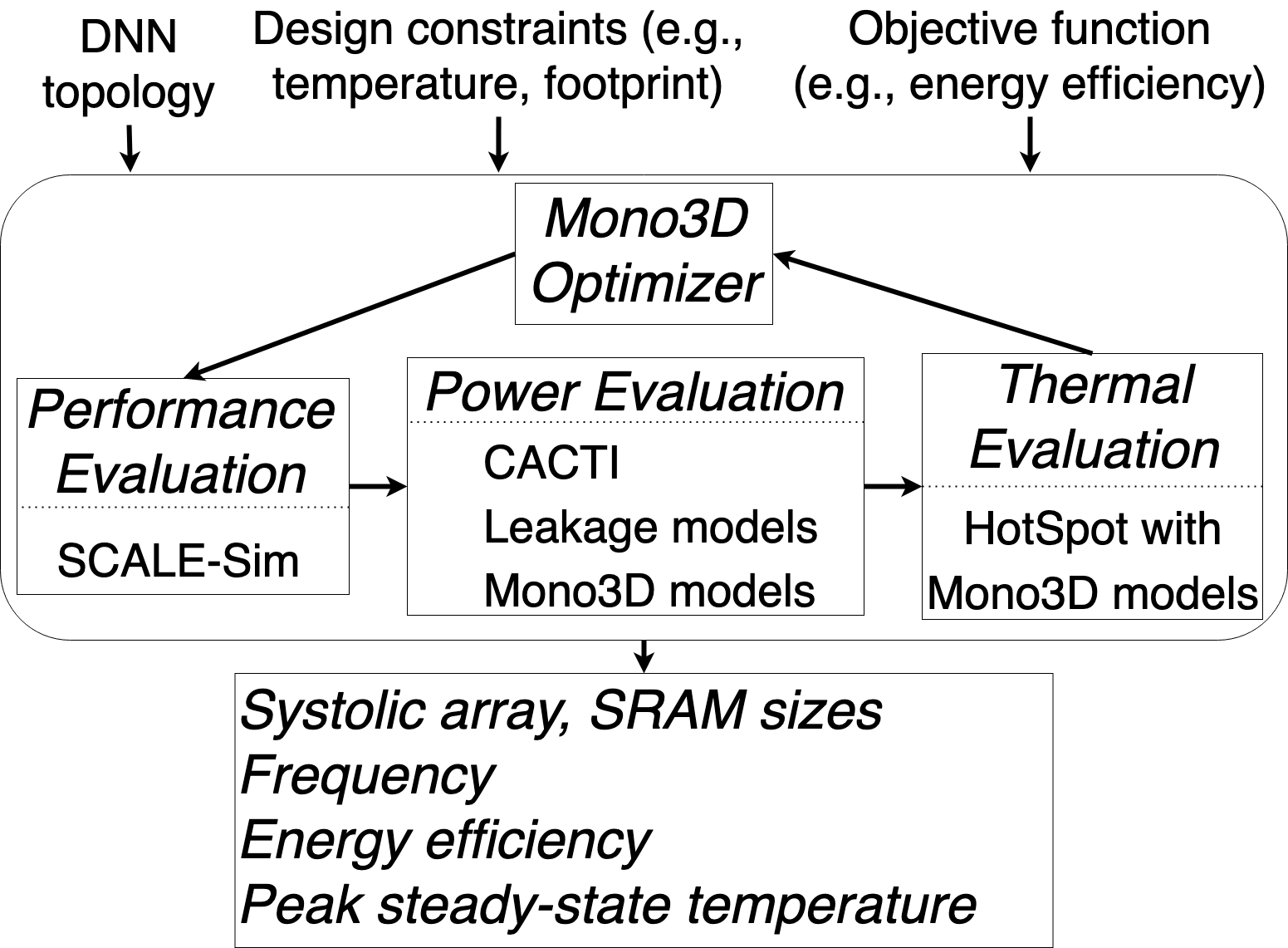}
	}
	\subfloat[4$\times$4 systolic array.]{
	\label{fig:systolic}
	\includegraphics[trim= 0mm 0cm 0mm 0mm,clip,height=3.5cm,width=.3\columnwidth,keepaspectratio]{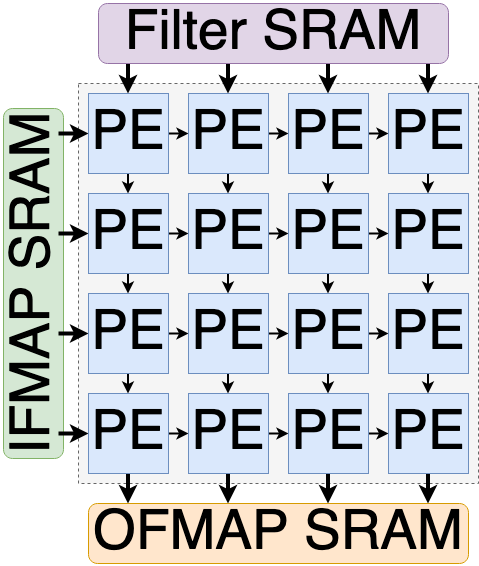}
	}
	\caption{\technologyname{} optimization flow and a sample systolic array.}
\end{figure}

We show a temperature-aware optimization flow in Fig. \ref{fig:optimizer} \cite{shukla2021temperature}. 
\begin{figure*}[htb!]
	\centering
	\vspace{-1.5em}
	\subfloat[\technologyname{} chipstack.]{
	\label{fig:chipstack}
	\includegraphics[trim= 0mm 0cm 0mm 0mm,clip,height=3cm,width=0.3\columnwidth,keepaspectratio]{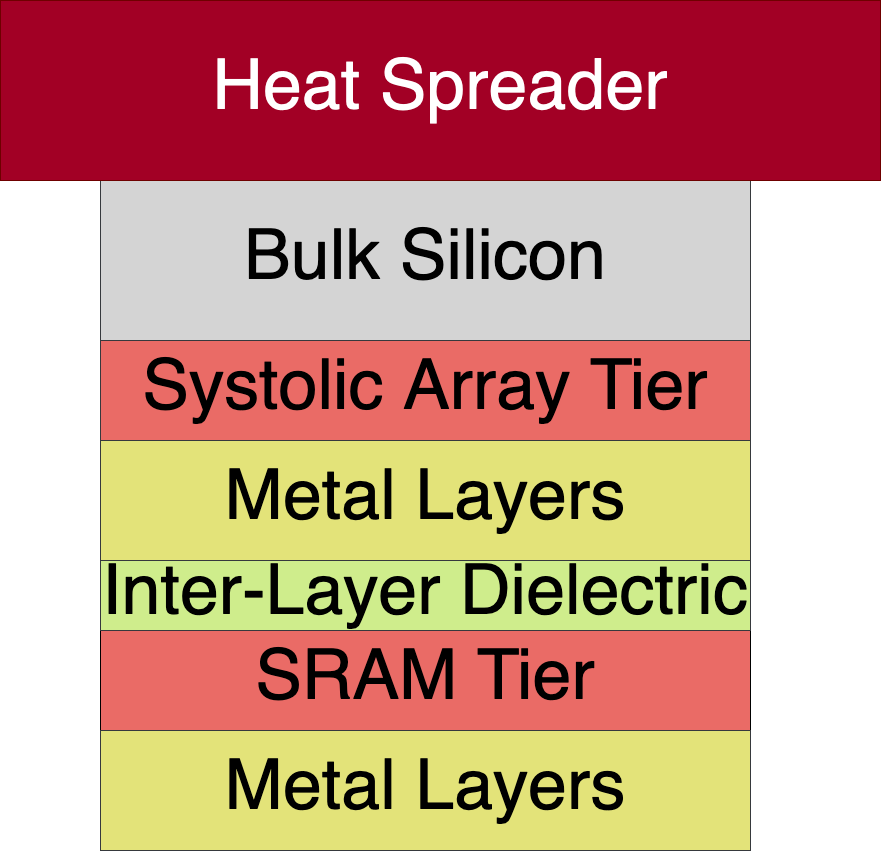}
	}
	\subfloat[\resnet{}]{
	\label{fig:Resnet50_Perf_Temp}
	\includegraphics[trim= 0mm 0mm 0mm 0mm,clip,height=3cm,width=0.6\columnwidth]{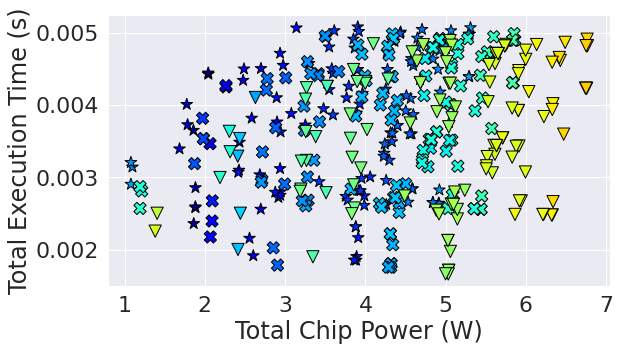}
    }
	\subfloat[\unet{}]{
	\label{fig:unet_Perf_Temp}
	\includegraphics[trim= 0mm 0mm 0mm 0mm,clip,height=3cm,width=0.7\columnwidth]{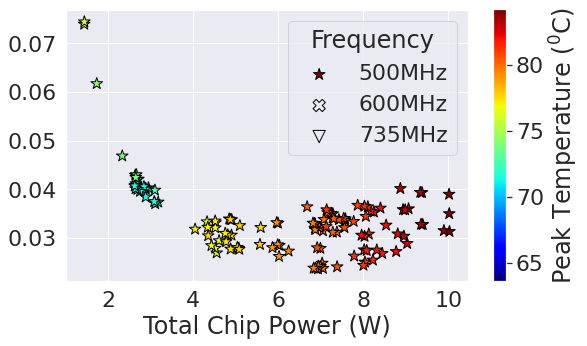}
    }
	\caption{Cross-sectional view of \technologyname{} chipstack (left) and performance versus power tradeoffs in \technologyname{} DNN accelerators.}
	\label{fig:results}
\end{figure*}
The inputs to the optimizer are design constraints (latency, temperature, footprint), a DNN and its topology (input/filter size, number of filters/channels, etc.), \blue{and} an objective function (energy efficiency). A multi-start simulated annealer (MSA)-based \orange{optimizer iterates} through performance, power, and thermal evaluation and converges to a near-optimal \technologyname{} configuration with safe chip temperature when it can no longer find better configurations. Multiple starts in MSA \blue{increase} the probability of escaping local optima and converging to global optima. We show our target DNN accelerator, a systolic array, in Fig. \ref{fig:systolic}. Systolic arrays are a 2D network of processing elements (PEs) with SRAMs for input feature map (IFMAP), weights (Filter), and outputs (OFMAP). Each PE is a multiply-and-accumulate (MAC) unit with internal registers for inputs/partial sums. Inputs are read from the top and left edges and passed on to the PEs in every clock cycle (Fig. \ref{fig:systolic}).

Several tools and models are integrated into the optimizer to evaluate \technologyname{} systolic array configurations. As shown in Fig. \ref{fig:optimizer}, the optimizer starts with performance evaluation of the DNN using SCALE-Sim, a cycle-accurate stall-free DNN inference simulator for systolic arrays \cite{samajdar2020systematic}, followed by power evaluation using CACTI-6.5 \cite{thoziyoor2009cacti} and \technologyname{} power models. For thermal evaluation, the optimizer uses HotSpot-6.0 to obtain steady-state temperatures \cite{skadron2003temperature}. There also exists a leakage-temperature loop for an accurate power/temperature estimation. The loop converges when the difference between consecutive HotSpot simulations is $<$ 1$^\circ$C.

We investigate a \technologyname{} configuration comprising two tiers, as shown in Fig. \ref{fig:chipstack}. The logic layer, i.e., systolic array tier is closer to the heat spreader \magenta{and} MIVs are used for SRAM read/writes. For simplicity, we assume that the systolic array and SRAM tiers are roughly equal in size \cite{shukla2021temperature}. We adopt a representative \technologyname{} power model for interconnect power from a recent work  \cite{shukla2021temperature}. A simplifying assumption made in this power model is that the interconnect power equals 15\% of the total chip dynamic power. On top of this interconnect power, 10\% interconnect power savings are applied for \technologyname{} power savings at iso-performance \cite{shukla2021temperature}. 
We also adopt a representative thermal model from a recent work \cite{yan2017mono3d} composed of metal layers, dielectric, etc. with the corresponding layer thicknesses and thermal resistivities.


\section{Experimental Results}
\label{sec:results}
To demonstrate the benefits of thermal awareness in the design of DNN systolic arrays for biomedical applications, we use two DNNs: \unet{} and \resnet{}. \cyan{Table \ref{tab:design_space} shows our design space}. We set a thermal threshold \pink{of} 80$^\circ$C \cyan{and} a limit on maximum performance loss of $\leq$ 10\% with respect to a latency-optimized configuration.
We use \pink{an example} MAC unit's area, power, and frequency, and include three frequency levels in our analysis: (500, 600, 735) MHz \cite{shukla2021temperature}. In total, there are 6k unique configurations for each DNN, including the frequencies. We launch six starts for each frequency with five perturbations. MSA parameters are set to 
\blue{1.44/0.88,}
0.85 for initial/final annealing temperatures\footnote{Annealing temperature: Unitless MSA parameter to determine when to accept a worse solution. Rate of cooling: Decaying rate of the annealing temperature to achieve convergence.} and rate of cooling, respectively \cite{shukla2021temperature}. 

Table \ref{tab:results} lists the \pink{optimized configurations} for inference latency, chip power, and energy-delay-area product (EDAP). We utilize EDAP to measure energy- and area-efficiency. Figures \ref{fig:Resnet50_Perf_Temp} and \ref{fig:unet_Perf_Temp} show the configurations explored by the optimizer for power minimization. Absence of a frequency level depicts that the optimizer did not find a valid configuration for initialization \pink{at that frequency}. As shown in the table, the optimizer converges to lowest frequency level for \unet{} and highest frequency level for \resnet{}. This difference is due to the topological differences among these DNNs. \resnet{} downsizes the input to make a final prediction for object classification, which leads to lower systolic array 
utilization, lower power, and fewer thermal violations (Fig. \ref{fig:Resnet50_Perf_Temp}). On the other hand, \unet{} first downsizes and then expands the input to obtain a high image resolution.
 Due to a larger input size in its latter layers, the array utilization is greater than in \resnet{}, thus leading to higher power and more thermal violations (\blue{Fig.} \ref{fig:unet_Perf_Temp}). The table also shows that due to the imposed constraint in performance loss, the optimizer converges to $\approx$53\% larger systolic arrays for  \unet{} at 500 MHz than \resnet{} at 735 MHz. In comparison to latency-optimized configuration, the power- and EDAP-optimized configurations achieve 21\% and 61\% improvement in chip power and EDAP, respectively, while sacrificing only 9.5\% in latency for \unet{}. \resnet{} achieves 49\% and 83\% improvement in chip power and EDAP using the optimizer, while sacrificing only 7.25\% in latency. 
We also compare \pink{these results with} unoptimized points corresponding to the smallest configuration in our design space (64$\times$68 with 352 $KB$ SRAM) running at the lowest frequency of 500 MHz, thus characterized by low power and area. Even though these configurations have lower power (avg. 55\%), the latencies are 3$\times$ (\unet{}) and 2$\times$ (\resnet{}) of the fastest configurations due to fewer PEs. While the unoptimized configuration has 35\% lower EDAP  for \resnet{}, for \unet{} this results in 50\% higher EDAP due to longer latency.
The above results show the importance of temperature-awareness in optimizing DNN accelerators for different objectives and DNNs. In addition, it \blue{motivates the need for} \magenta{systematic} optimization \blue{to balance constraints and objectives in a thermally-aware manner}.


\begin{table}[htb!]
    \centering
    \resizebox{0.8\columnwidth}{!}{
    \begin{tabular}{|c|c|} \hline
        Systolic array size & 64$\times$64 to 256$\times$256  \\ \hline
        Each SRAM size & (32, 64 ... 4096) $KB$\\ \hline
        Aspect ratio of the chip &  0.94 to 1 \\ \hline
        Frequencies &  (735, 600, 500) MHz \\ \hline
    \end{tabular}}
    \caption{Design space for DNN accelerators.}
    
    \label{tab:design_space}
    \vspace{-0.1in}
    \vspace{-0.05in}
\end{table}

\begin{table}[htb!]
\resizebox{\columnwidth}{!}{
\vspace{-.1em}
\begin{tabular}{|c|c|c|c|c|c|c|c|c|}
\hline
 
\textit{\textbf{Optimization Goal}} & \unet{} & \resnet{} \\ \hline
\begin{tabular}[c]{@{}c@{}}Performance\\ (Inference Latency)\end{tabular} &
\begin{tabular}[c]{@{}c@{}}194$\times$192 (500 MHz)\\ 4256 $KB$\end{tabular} & 
\begin{tabular}[c]{@{}c@{}}186$\times$196 (735 MHz)\\ 4160 $KB$\end{tabular} \\ \hline
Chip Power & 
\begin{tabular}[c]{@{}c@{}}162$\times$172 (500 MHz)\\ 3136 $KB$\end{tabular} & 
\begin{tabular}[c]{@{}c@{}}132$\times$138 (735 MHz) \\ 2112 $KB$\end{tabular} \\ \hline
System EDAP & 
\begin{tabular}[c]{@{}c@{}}162$\times$172 (500 MHz)\\ 3136 $KB$\end{tabular} & 
\begin{tabular}[c]{@{}c@{}}134$\times$136 (735 MHz)\\ 2112 $KB$\end{tabular} \\ \hline
\end{tabular}
}
\caption{Optimization results: Systolic array (operating frequency) and total SRAM (IFMAP, Filter, OFMAP).}
\label{tab:results}
\end{table}
\section{Conclusion}
\label{sec:conclusion}

We demonstrate the effectiveness of including temperature-awareness in design optimization for \technologyname{} energy efficient DNN accelerators, subject to user-defined performance and thermal constraints for biomedical applications. Since \unet{} dissipates high power and results in higher temperature, the optimizer converges to \technologyname{} configurations \magenta{operating} at \magenta{a} lower frequency for energy efficiency. For \resnet{}, the optimizer utilizes the thermal slack and converges to configurations \magenta{operating} at \magenta{a} higher frequency due to fewer thermal violations.

\bibliographystyle{IEEEtran}
{\footnotesize\bibliography{IEEEabrv,ref.bib}}

\begin{thebibliography}{10}
\providecommand{\url}[1]{#1}
\csname url@samestyle\endcsname
\providecommand{\newblock}{\relax}
\providecommand{\bibinfo}[2]{#2}
\providecommand{\BIBentrySTDinterwordspacing}{\spaceskip=0pt\relax}
\providecommand{\BIBentryALTinterwordstretchfactor}{4}
\providecommand{\BIBentryALTinterwordspacing}{\spaceskip=\fontdimen2\font plus
\BIBentryALTinterwordstretchfactor\fontdimen3\font minus
  \fontdimen4\font\relax}
\providecommand{\BIBforeignlanguage}[2]{{%
\expandafter\ifx\csname l@#1\endcsname\relax
\typeout{** WARNING: IEEEtran.bst: No hyphenation pattern has been}%
\typeout{** loaded for the language `#1'. Using the pattern for}%
\typeout{** the default language instead.}%
\else
\language=\csname l@#1\endcsname
\fi
#2}}
\providecommand{\BIBdecl}{\relax}
\BIBdecl

\bibitem{ronneberger2015u}
O.~Ronneberger \emph{et~al.}, ``{U-net: Convolutional networks for biomedical
  image segmentation},'' in \emph{MICCAI}.\hskip 1em plus 0.5em minus
  0.4em\relax Springer, 2015, pp. 234--241.

\bibitem{al2020breast}
Q.~A. Al-Haija and A.~Adebanjo, ``Breast cancer diagnosis in histopathological
  images using {R}esnet-50 convolutional neural network,'' in
  \emph{IEMTRONICS}.\hskip 1em plus 0.5em minus 0.4em\relax IEEE, 2020, pp.
  1--7.

\bibitem{wei2020review}
Y.~Wei \emph{et~al.}, ``A review of algorithm \& hardware design for {AI}-based
  biomedical applications,'' \emph{IEEE TBioCAS}, vol.~14, no.~2, 2020.

\bibitem{batude20143d}
P.~Batude \emph{et~al.}, ``{3D sequential integration opportunities and
  technology optimization},'' in \emph{IEEE Int. Interconnect Tech. Conf.},
  2014, pp. 373--376.

\bibitem{hu2018stacking}
X.~Hu \emph{et~al.}, ``Die stacking is happening,'' \emph{IEEE Micro '18},
  vol.~38, no.~1, pp. 22--28, 2018.

\bibitem{shukla2019overview}
P.~Shukla \emph{et~al.}, ``{An overview of thermal challenges and opportunities
  for monolithic 3D ICs},'' 2019, pp. 439--444.

\bibitem{kung1982systolic}
H.-T. Kung, ``Why systolic architectures?'' \emph{Computer}, no.~1, pp. 37--46,
  1982.

\bibitem{shukla2021temperature}
P.~Shukla \emph{et~al.}, ``Temperature-aware optimization of monolithic 3{D}
  deep neural network accelerators,'' in \emph{IEEE ASP-DAC}, 2021, pp.
  709--714.

\bibitem{samajdar2020systematic}
A.~Samajdar \emph{et~al.}, ``A systematic methodology for characterizing
  scalability of {DNN} accelerators using {SCALE-S}im,'' in \emph{IEEE ISPASS},
  2020.

\bibitem{thoziyoor2009cacti}
S.~Thoziyoor \emph{et~al.}, ``{CACTI} 6.5,'' \emph{hpl.hp.com}, 2009.

\bibitem{skadron2003temperature}
K.~Skadron \emph{et~al.}, ``Temperature-aware microarchitecture,'' \emph{ACM
  SIGARCH Computer Architecture News}, vol.~31, no.~2, pp. 2--13, 2003.

\bibitem{yan2017mono3d}
C.~Yan and E.~Salman, ``{Mono3D: Open source cell library for monolithic 3-D
  integrated circuits},'' \emph{IEEE TCAS I}, vol.~65, no.~3, 2018.

\end{thebibliography}

\end{document}